\documentclass[twocolumn,showpacs,preprintnumbers,amsmath,amssymb]{revtex4}
\usepackage{graphicx}
\usepackage{rotating}
\begin{document}
\title{Coulomb and nuclear breakup of a halo nucleus $^{11}$Be}
\author{N.~Fukuda$^1$, T.~Nakamura$^{2}$, 
N.~Aoi$^1$, N.~Imai$^1$, M.~Ishihara$^{1}$, T.~Kobayashi$^3$ 
H.~Iwasaki$^4$, T.~Kubo$^1$, A.~Mengoni$^{1,5}$, M.~Notani$^6$, 
H.~Otsu$^3$,
H.~Sakurai$^4$, S.~Shimoura$^6$, T.~Teranishi$^6$, 
Y.X.~Watanabe$^1$, K.~Yoneda$^1$}
\affiliation{
$^1$ RIKEN, 2-1 Hirosawa, Wako,
Saitama 351-0198, Japan\\
$^2$ Department of Physics, Tokyo Institute of Technology, 2-12-1
O-Okayama, Meguro, Tokyo 152-8551, Japan\\
$^3$ Department of Physics, Tohoku University, 2-1 Aoba, Aramaki,
Aoba, Sendai 980-8578, Japan\\
$^4$ Department of Physics, University of Tokyo, 7-3-1 Hongo, Bunkyo, Tokyo
113-0033, Japan\\
$^5$ Applied Physics Division, ENEA, Bologna 2, I-40129, Italy\\
$^6$ CNS, University of Tokyo, RIKEN campus, Hirosawa 2-1, Wako, Saitama
351-0198, Japan\\
}
\date{\today}

\newcommand{\dsde}{\frac{d\sigma_{\rm CD}}{dE_{\rm x}}}
\newcommand{\dsderel}{\frac{d\sigma_{\rm CD}}{dE_{\rm rel}}}
\newcommand{\dsdenf}{d\sigma_{\rm CD}/dE_{\rm x}}
\newcommand{\dbde}{\frac{dB({\rm E1})}{dE_{\rm x}}}
\newcommand{\dbderel}{\frac{dB({\rm E1})}{dE_{\rm rel}}}
\newcommand{\dbdenf}{dB({\rm E1})/dE_{\rm x}}
\newcommand{\dbderelnf}{dB({\rm E1})/dE_{\rm rel}}
\newcommand{\neone}{N_{\rm E1}(\theta,E_{\rm x})}
\newcommand{\erel}{E_{\rm rel}}
\newcommand{\ex}{E_{\rm x}}
\newcommand{\jpi}{J^\pi}
\newcommand{\shalfzero}{^{10}{\rm Be(0}^+{\rm )}\otimes \nu 2s_{1/2}}
\newcommand{\shalftwo}{^{10}{\rm Be(2}^+{\rm )}\otimes \nu 2s_{1/2}}
\newcommand{\dhalfzero}{^{10}{\rm Be(0}^+{\rm )}\otimes \nu 1d_{5/2}}
\newcommand{\dhalftwo}{^{10}{\rm Be(2}^+{\rm )}\otimes \nu 1d_{5/2}}
\newcommand{\beone}{B{\rm (E1)}}
\newcommand{\thetad}{\theta_{\rm D}}
\newcommand{\dsdomegade}{\frac{d\sigma^2}{d\Omega d\erel}}
\newcommand{\dsdomega}{\frac{d\sigma(\theta,\ex)}{d\Omega}}
\newcommand{\dndomega}{\frac{dN_{\rm E1}(\theta,\ex)}{d\Omega}}
\newcommand{\dndomeganf}{dN_{\rm E1}/d\Omega}
\newcommand{\beoneex}{B{\rm (E1;}E_{\rm x})}
\newcommand{\beonedex}{\frac{d\beone}{d\ex}}
\newcommand{\beonedrel}{\frac{d\beone}{d\erel}}
\newcommand{\massbe}{M(^{11}{\rm Be}^{*})}
\newcommand{\vecbeam}{\mbox{\boldmath$P$}(^{11}{\rm Be})}
\newcommand{\vecfrag}{\mbox{\boldmath$P$}(^{10}{\rm Be})}
\newcommand{\vecneut}{\mbox{\boldmath$P$}(n)}
\newcommand{\efrag}{E(^{10}{\rm Be})}
\newcommand{\eneut}{E(n)}
\newcommand{\mfrag}{m(^{10}{\rm Be})}
\newcommand{\mneut}{m(n)}
\newcommand{\zerosix}{0^\circ~\leq\theta\leq~6^\circ}
\newcommand{\zerotwelve}{0^\circ~\leq~\theta\leq~12^\circ}
\newcommand{\zeroonethree}{0^\circ~\leq\theta\leq~1.3^\circ}
\newcommand{\zerozerofive}{0^\circ~\leq\theta~\leq~0.5^\circ}
\newcommand{\zeroone}{0^\circ~\leq\theta~\leq~1^\circ}
\newcommand{\threesix}{3^\circ~\leq\theta~\leq~6^\circ}
\begin{abstract}
Breakup reactions of the one-neutron halo nucleus 
$^{11}$Be on lead and carbon targets at about 70~MeV/nucleon have been
investigated at RIKEN by measuring the momentum vectors of the 
incident $^{11}$Be, outgoing $^{10}$Be, and neutron in coincidence.
The relative energy spectra as well as the angular
distributions of the $^{10}$Be+n center of mass system (inelastic angular
distributions) have been extracted both for Pb and C targets. 
For the breakup of $^{11}$Be on Pb, 
the selection of forward scattering angles,
corresponding to large impact parameters,
is found to be effective to extract almost purely 
the first-order E1 Coulomb breakup
component, and to exclude the nuclear contribution and higher-order
Coulomb breakup components. This angle-selected energy spectrum is thus
used to deduce the spectroscopic factor for
the $\shalfzero$ configuration in $^{11}$Be which is found to be 
0.72~$\pm$~0.04 with a $\beone$ strength up to $\ex$~=~4~MeV of
1.05~$\pm$~0.06~e$^2$fm$^2$. The energy weighted E1 strength
up to $\ex$~=~4~MeV explains 70~$\pm$~10~\% of the cluster sum rule, 
consistent with the obtained spectroscopic factor.
The non-energy weighted sum rule within the same energy range 
is used to extract the
root mean square distance of the halo neutron
to be 5.77(16)~fm, consistent with previously known values.
In the breakup with the carbon target, we have 
observed the excitations to the known unbound states in $^{11}$Be 
at $\ex$~=~1.78~MeV and $\ex$~=~3.41~MeV. 
Angular distributions for these states show the diffraction 
pattern characteristic of $L$=2 transitions, 
resulting in $J^{\pi}$=(3/2,5/2)$^+$
assignment for these states. 
We finally find that even for the C target
the E1 Coulomb direct breakup mechanism becomes dominant 
at very forward angles.

\end{abstract}
\maketitle


\section{Introduction}

Breakup reactions have played key roles in investigating
the properties of weakly-bound halo nuclei over the past
decade~\cite{HANS95,TANI95}. 
The breakup reaction on a light target, induced predominantly 
by the nuclear interaction, is characterized by an
unusually narrow momentum distribution of a 
core fragment and an enhanced reaction cross section, reflecting the
extended neutron halo structure.
Indeed, the halo structure
was first uncovered for $^{11}$Li by observing the enhanced interaction
cross section for this nucleus~\cite{TANI85}, and the narrow momentum 
distribution of $^9$Li following the breakup of $^{11}$Li 
on a carbon target~\cite{KOBA88,ORR92}.
In addition to these techniques, more recently, the
one nucleon knockout reaction in coincidence with $\gamma$ rays from
the core fragment has been successfully used to determine
spectroscopic factors of halo states~\cite{AUMA00}.

The breakup reaction of halo nuclei on a heavy target
predominantly occurs as Coulomb breakup (Coulomb dissociation).
This reaction is of particular interest due to 
substantially enhanced Coulomb breakup cross sections
found for halo nuclei~\cite{KOBA89}.
This phenomenon was first interpreted as the presence
of a soft electric dipole (E1) resonance~\cite{IKED92},
which occurs as a vibration of the core against halo due to the 
low density of the halo cloud. 
More recently, by using kinematically complete measurements
of the Coulomb breakup, spectra of electric dipole strength ($\beone$)
have been directly obtained for one-neutron halo nuclei
$^{11}$Be~\cite{NAKA94} and $^{19}$C~\cite{NAKA99}, 
and two-neutron halo nuclei $^{6}$He~\cite{AUMA99}, 
$^{11}$Li~\cite{IEKI93,SHIM95,ZINS97}, and $^{14}$Be~\cite{LABI01}. 
It was found that for these halo nuclei a strong E1 strength
of the order of 1~W.u.(Weisskopf Unit) 
was observed at very low excitation energies of about 1~MeV.
However, the mechanism for
such large E1 strength was not due to a soft dipole resonance, 
but rather due to a direct breakup into the continuum, as shown 
by our earlier study of the  Coulomb breakup of $^{11}$Be~\cite{NAKA94}. 
In the direct breakup mechanism, the observed enhancement of the
E1 strength is interpreted as follows: 
the $\beone$ distribution is described approximately as a  
Fourier transform of $rR(r)$, where $r$ is the radial coordinate of
the neutron and $R(r)$ the radial component of the 
wave function of the halo neutron~\cite{OTSU94}.
The large value of $|R(r)|^2$ at large
$r$ in a halo nucleus thus leads to a large E1 strength 
at low excitation energies. In fact, the $\beone$ distribution can be
used to determine $R(r)$ by inverse Fourier
transformation~\cite{NAKA94,MENG97}.

In this article, we will show the results of a new,
full kinematical measurement of the
breakup reactions of $^{11}$Be with a heavy target (lead) where Coulomb 
breakup dominates, and with a light target (carbon) where nuclear breakup
dominates. We aim at a comprehensive understanding of the reaction mechanism
of the breakup reactions both on heavy and light targets, thereby 
establishing a way of doing spectroscopy of halo nuclei by
the breakup reactions, both for the 
ground state as well as for excited states in the continuum.

For the breakup with a Pb target, we
focus mainly on extracting the Coulomb breakup component 
by using the information on the scattering angle, which
is approximately inversely proportional to the impact parameter of 
the reaction. 
The analysis incorporating the scattering angle dependence
has been obtained with much more statistics (more than 30 times) as
compared to the previous experiment~\cite{NAKA94}. The contribution of
the nuclear breakup component and higher order effects in the breakup
with a heavy target have recently drawn much attention. 
In fact, quite a few theoretical papers have suggested 
the necessity of careful treatments of these 
contributions~\cite{NAGA01,DASS99,TYPE01a,TYPE01b,BERT92,
BAUR92,ESBE95,MARG02,MARG03,THOM01,BAUR03} beyond the direct
breakup mechanism based on a semi-classical first-order perturbation
theory (equivalent photon method) 
which we successfully applied in the analysis of 
the previous experiment~\cite{NAKA94,NAKA99}. 
For instance, a much larger nuclear contribution
than the conventional estimation made by
scaling the breakup cross section with the data obtained with a light
target
has been suggested in Ref.~\cite{NAGA01,DASS99}.
In this article, we prove that the direct breakup mechanism
with the first-order Coulomb breakup is dominant, 
and the small nuclear contribution and higher order effects
can be well controlled using the angular distribution of the
center-of-mass system of $^{10}$Be~+~$n$.
This technique will thus offer a powerful spectroscopic tool 
that can extract precisely the halo wave function $R(r)$.

For the breakup with a C target, we focus on investigating the
excitation of discrete resonant states by using the information of 
the excitation energy spectrum 
in combination with the scattering angle.
Thereby, we aim at establishing a spectroscopic method to 
study the narrow discrete states in the continuum. 
Such states are hardly observed in the breakup 
with a heavy target due to the large direct-breakup contribution.
We also examine the structureless part of the energy spectrum with the
scattering angle distribution which is used to disentangle the
reaction mechanism with the light target.

The $^{11}$Be nucleus is a suitable test case
for these studies since ground state properties are well known.
For example,
the one-neutron separation energy $S_n$ is precisely known to be
504~$\pm$~4 keV~\cite{AUDI93}. Furthermore, the simple one-neutron
halo structure of $^{11}$Be has an advantage over two-neutron halo nuclei
such as $^{11}$Li because the reaction mechanisms do not 
suffer from the complexity
which may arise from the two-neutron halo correlations.

The breakup reactions of $^{11}$Be on targets from light to 
heavy mass have been studied by R.~Anne {\it et al.}~\cite{ANNE93}. 
The authors used mainly the inclusive neutron angular distributions.
More recently, breakup reactions for $^{11}$Be have been studied at GSI
in a full-kinematical way using Pb and C targets and at high energy, 
520~MeV/nucleon~\cite{PALI03}. 
Our present approach is a full-kinematical one.
We can extract the excitation energy spectrum as well as the scattering
angle of the c.m.~(center of mass) system of $^{10}$Be and neutron.
In particular, in this paper
we emphasize the importance of the information on the 
scattering angle, which was not discussed
in the GSI data. In addition, we have performed the
experiment at a much lower energy as compared to the one at GSI,
bringing in additional information on the reaction mechanism.

We organize the paper as follows: Section~\ref{method} 
describes the experimental method. 
Section~\ref{setup} describes the experimental setup. 
In Sec.~\ref{results} the results for breakup of $^{11}$Be on
Pb and C targets are presented with detailed discussions including 
theoretical comparisons.
Then, in Sec.~\ref{summary}, the conclusions are given.

\section{EXPERIMENTAL METHOD}
\label{method}

In the current experiment, we made a coincidence measurement of
the momentum vectors of the incoming $^{11}$Be, outgoing $^{10}$Be, and 
neutron to deduce
the relative energy $\erel$ between $^{10}$Be and the neutron, 
and the scattering angle $\theta$ of the c.m. system of $^{10}$Be~+~n. 
Here we describe the features characteristic of the invariant mass
method which has been used to extract $\erel$, and the method to
extract $\theta$.

\subsection{Invariant mass method}

The relative energy $\erel$ between $^{10}$Be and the neutron, which
is related to the excitation energy $\ex$ of $^{11}$Be 
by $\erel=\ex -S_n$,
can be extracted by using the
invariant mass method. The invariant mass $\massbe$ of 
the intermediate excited state of $^{11}$Be is 
determined by measuring the momentum vectors,
$\vecfrag$ and $\vecneut$, of the outgoing particles $^{10}$Be and
neutron, respectively. 
Namely,
\begin{equation}
\massbe = \sqrt{(\efrag + \eneut)^2 - (\vecfrag + \vecneut)^2},
\end{equation}
where $\efrag$ and $\eneut$ stand for the total energy of the $^{10}$Be
fragment and the neutron, respectively. 
The relative energy $\erel$ between $^{10}$Be and the neutron is then
determined as,
\begin{equation}
\erel = \massbe - \mfrag - \mneut,
\end{equation}
where $\mfrag$ and $\mneut$ denote the mass of $^{10}$Be and of the neutron, 
respectively.

The advantage of the invariant mass method
is that the energy resolution is as good as about 
a few hundred keV at $\erel$~=~1~MeV. 
This is due to the fact that the invariant mass is a function of
four momenta of the outgoing particles, and is not affected by the 
widely spread secondary beam. In this sense, this method is
appropriate for radioactive beam experiments.
The good energy resolution is also attributed to
the fact that $\erel$ is determined by the 
opening angle and the relative velocity between 
the outgoing $^{10}$Be and the neutron.
In this case, the opening angle 
resolution of 10~mrad and the relative velocity resolution of 1\%, which are
easily achievable, can yield a good 
energy resolution of a few hundred keV at $\erel$~=~1~MeV.
This is different from the missing mass method, where the
resolution is determined by the value of the total
mass which is of the order of tens of GeV. Thus, an energy 
resolution of the order of 10~MeV, even with the momentum 
resolution of the order of 0.1\%, can only be achieved.
Further advantages are the kinematical focusing and the
availability of a thick target since the projectile has rather
high velocity of more than 0.3$c$ for intermediate incident 
energies. Relatively small detectors can thus cover most of the
acceptance, which is very 
important in radioactive beam experiments as well.

It should be noted that there is a possibility that the 
$^{10}$Be fragment is produced in an excited state. 
In this case, a $\gamma$ ray emission follows the reaction process
and has been measured in the GSI experiment~\cite{PALI03}. 
The excitation energy in this case has to be modified to 
$\ex$~=~$\erel$~+~$S_n$~+~$E_\gamma$, where $E_\gamma$ stands for the
energy of the deexcitation $\gamma$ ray from the daughter
nucleus $^{10}$Be.
In the current work, we did not use $\gamma$
ray detectors. However, the probability of obtaining an
excited $^{10}$Be, where
the lowest excited state is located as high as 3.37~MeV, 
is very small for the Coulomb breakup process
due to very small virtual photon numbers
for this high excitation energy. Since the ratio of
the virtual photon number at higher $\ex$ to that at lower 
$\ex$ is smaller for lower incident energies, this probability is
even smaller in our case. In fact, 
this contribution is estimated to be less than 3\% at the present 
incident energy of 69~MeV/nucleon compared to 6\% observed at 
520~MeV/nucleon~\cite{PALI03}.
The selection of large impact parameters done 
in the current analysis further reduces this number,
leading to a negligible contribution 
resulting from the excited $^{10}$Be
states. For the breakup with a light target, the GSI experiment found
about 17\% contribution of the non ground state component, and thus
the treatment required additional care.

\subsection{Scattering angle of the center of mass}

The exclusive measurement of an
incident $^{11}$Be momentum $\vecbeam$ in addition to 
$\vecfrag$ and $\vecneut$ allowed us to extract 
the scattering angle $\theta$ of the c.m. system of $^{10}$Be+n.
This angle is determined by the opening angle between 
the direction of $\vecbeam$ and that of 
the outgoing momentum vector of the center of mass
obtained by $\vecfrag+\vecneut$. 
Here, the scattering angle $\theta$ is defined
in the center of mass frame of the projectile and target. 

Since we are dealing with a small relative energy of 
less than 5~MeV compared to the total kinetic energy of 
about 770~MeV, the angle determined in this way represents the
inelastic scattering angle of $^{11}$Be on the Pb or C target
with very good approximation.
For the Coulomb breakup, the scattering angle 
is directly related to the impact parameter as will be shown
for the semi-classical approximation.
For the nuclear breakup, the scattering angle is used mainly to 
determine the orbital angular momentum transfer 
$L$ in the transition to a given discrete state, and thus can be used to
assign the spin-parity $J^\pi$ of the excited state.

\section{Experiments}
\label{setup}

\begin{figure*}[t]
\includegraphics[width=140mm]{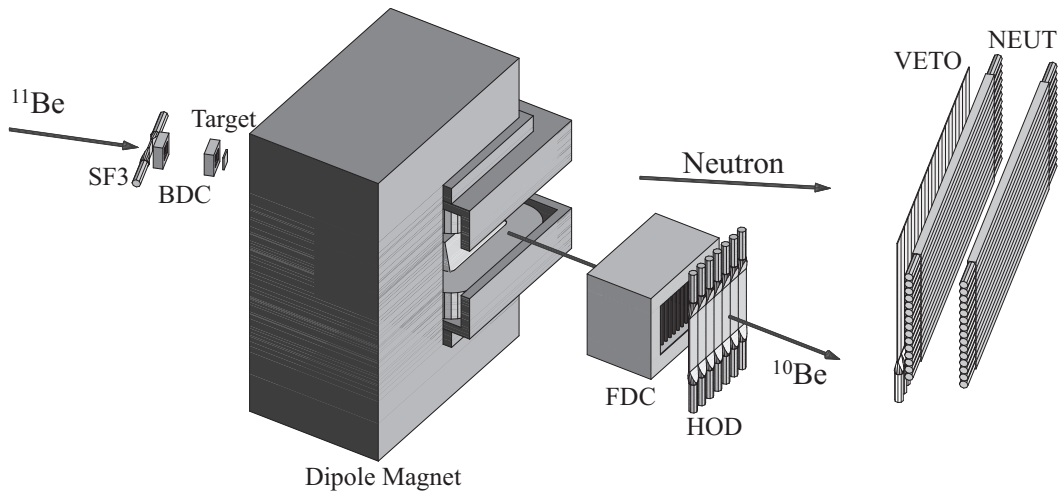}%
\caption{\label{fig_setup} A schematic view of the experimental setup 
located at the last focal point (F3) of RIPS. 
The setup consists of a beam scintillator~(SF3), tracking drift chambers
for the secondary beam particle~(BDC), 
a neutron detector array~(NEUT), charged particle veto
detectors~(VETO), a dipole magnet and an associated
drift chamber for a charged fragment~(FDC), 
and a charged particle hodoscope~(HOD).}
\end{figure*}

The experiment was performed at the RIKEN Accelerator Research Facility
(RARF). A secondary radioactive beam of $^{11}$Be 
was produced by fragmentation of a $^{18}$O 
primary beam at 100~MeV/nucleon in a thick Be target. 
The secondary beam was separated using the RIPS 
fragment separator~\cite{KUBO92}, where an achromatic
wedge-shaped energy degrader was installed at the intermediate 
dispersive focal plane to adjust the secondary beam energies to
about 70~MeV/nucleon and to purify the secondary beam.  
The typical beam intensity was 
restricted to about 5$\times$10$^4$ particles/sec by setting 
the momentum slit down to $\Delta P/P\leq\pm$0.1\%  in order to meet the
counting capabilities of the detectors. 
The resulting $^{11}$Be beam with a purity of about 99\% 
was delivered to the experimental setup
shown schematically in Fig.~\ref{fig_setup}.

The $^{11}$Be ion bombarded a $^{nat}$Pb 
target with a thickness of 224 mg/cm$^2$ or
$^{nat}$C target with a thickness of 376~mg/cm$^2$.
In addition, a no-target run was performed to subtract the 
background events generated by materials other than the target.
The energy of the incident $^{11}$Be particle was determined from the time of
flight (TOF), measured with two thin plastic scintillators with a thickness
of 1~mm which were placed 4.57~m
apart along the beam line. The average beam energy at the 
mid-plane of the target was 68.7~MeV/nucleon and 67.0~MeV/nucleon,
respectively for the Pb and C targets. 
The position and angle of $^{11}$Be incident on the target 
were measured with
two sets of multi-wire drift chambers (BDC). 
The energy and angle of the incident particle
were combined to reconstruct the momentum vector of the projectile,
{\it i.e.}, $\vecbeam$.

The breakup particles, $^{10}$Be and $n$ were emitted in a narrow cone
at forward angles with velocities close to that of the $^{11}$Be
incident ion. The neutrons were detected by the two layers of a neutron
hodoscope array (NEUT), which has an active area of 
214(H)$\times$92(V)cm$^2$ and a depth of 6.1~cm for each layer. 
The front faces of NEUT were placed at 460 cm and 499 cm downstream of the 
target. The detector covered an angular range from $-7.0$ degrees
to $19.5$ degrees in the horizontal direction, and $\pm$5.6 degrees
in the vertical directions. 
NEUT consists of 30 plastic scintillator rods (15 rods for
each layer). Each detector has a dimension
of 6.1 cm(D)~$\times$~6.1 cm(V)~$\times$~214~cm(H), coupled to 
two photomultiplier tubes on both ends. The front side of
NEUT was equipped with a thin layer of plastic scintillators (VETO) 
set in order to reject the charged particle background.
The TOF information of the neutron was obtained by taking the mean value of the
two timings of the fired detector of NEUT. 
The horizontal position was
obtained by taking the difference of the two timings.
The vertical position was determined by the position of 
the fired rod. The momentum vector $\vecneut$ was thus reconstructed
from the position and TOF information of these detectors.
The momentum resolution (1$\sigma$) of the neutron 
in the projectile rest frame was 1.7\% and 2.0\% for the Pb and C
targets, respectively.
The intrinsic neutron detection efficiency of 13.4\% for the threshold energy
6~MeVee (electron-equivalent) was obtained from a separate
experiment using the $^7$Li($p,n$)$^7$Be reaction at 66.7~MeV.
This energy threshold was used to reject all the $\gamma$-ray related events.

The $^{10}$Be fragment emitted in the reaction 
was bent by a large-gap dipole
magnet, was traced by the multi-wire drift chamber (FDC) 
located downstream of the magnet, and penetrated the hodoscope (HOD) 
which consists of 7 plastic scintillator slats of 1~cm thickness.
Particle identification was performed by combining $\Delta E$ and TOF
information from the hodoscope with the magnetic rigidity information from
the tracking. 
The momentum vector of $^{10}$Be ($\vecfrag$) was 
deduced by the combination of TOF between the target and HOD (about 4~m),
and tracking analysis. 
The momentum resolutions (1$\sigma$) of
$^{10}$Be for the reaction with the Pb target 
were 0.80\%, 0.77\%, and 0.32\% respectively for the $P_x$,
$P_y$, and $P_z$, which represent the 
horizontal, vertical, and parallel momenta. Those for the C target
were 0.47\%, 0.47\%, and 0.32\%, respectively. This difference in the
energy resolution for the transverse directions according to the
target is due to the
different multiple scattering between the heavy and light targets.

The relative energy resolution was determined by a Monte-Carlo
simulation incorporating the 
momentum resolutions of $^{10}$Be and the neutron.
The relative energy resolution (FWHM) was thus estimated to be
$0.44\sqrt{\erel}$~MeV and $0.45\sqrt{\erel}$~MeV 
respectively for the Pb and
C targets. The angular resolution of $\theta$ in 1~$\sigma$ 
was 0.41 degrees and 0.48 degrees respectively for the
Pb and C targets. 

The geometrical acceptance for the $^{10}$Be and neutron was estimated by a
Monte Carlo simulation. Here, events were generated as a function
of $\erel$ and $\theta$, and the corresponding 
acceptance functions for the Pb and C targets  
were deduced for these observables. 
The net geometrical acceptance was obtained as a ratio of the breakup
events of interest with and without acceptance correction.
The acceptance thus estimated turned out to be 52\%
for the Pb target 
with the energy-angular ranges of 0~$\leq~\erel~\leq$~5~MeV and $\zerosix$ . 
The same quantity was 31\% for the C target, with the ranges of
0~$\leq~\erel~\leq$~8~MeV and $\zerotwelve$. 

\section{Results and Discussions}
\label{results}

\subsection{Overview of $\erel$ spectra for Pb and C targets}

\begin{figure}[t]
\includegraphics[width=85mm]{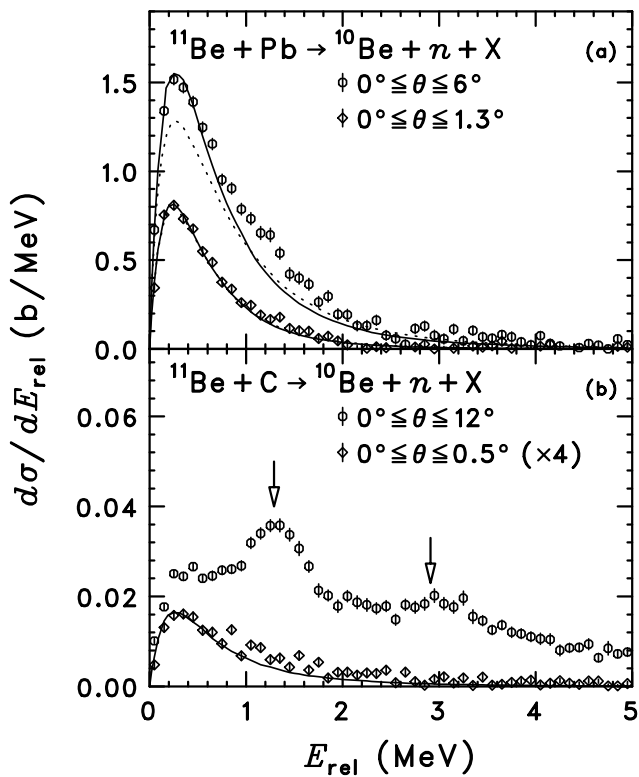}
\caption{
\label{erel} Relative energy spectra for $^{11}$Be+Pb at
69~MeV/nucleon (a), and for $^{11}$Be+C at 67~MeV/nucleon (b). 
These are plotted for the whole acceptance region (open circles), 
and for the selected forward angles (open diamonds). 
The data points are compared to the E1 direct breakup model calculation.
The solid curves are obtained with the ECIS code with 
$\alpha^2$ (spectroscopic factor for the halo configuration) 
of 0.72 (solid line), while the dotted curves are obtained with the
equivalent photon method with $\alpha^2$=0.69. 
For the carbon data, two discrete peaks 
corresponding to $\ex$~=~1.78~MeV and 3.41~MeV marked by the 
arrows are observed.}
\end{figure}

The relative energy spectra for the Pb target and C target data 
are shown in Figs.~\ref{erel}(a) and (b), respectively.
There, the cross sections for the breakup channel into
$^{10}$Be~+~$n$ are plotted for 
the angular range $\zerosix$ ($\zerotwelve$) corresponding to 
the current whole acceptance,
and for the selected forward angular
ranges $\zeroonethree$ ($\zerozerofive$) for the Pb(C) targets. 
The angular ranges for the whole acceptance are 
different depending on the target used because the angle $\theta$
in the projectile-target center of mass frame 
is about twice as much as that in the laboratory frame for the C target, 
while they are about the same for the Pb target.

The spectra for the whole acceptance show conspicuously different
characteristics depending on the target.
A huge asymmetric peak is seen for the Pb target, while two
peaks, corresponding to the known states at 
$\ex$~=~1.78~MeV and 3.41~MeV,
are seen on top of the decreasing continuum for the C
target.
The breakup cross sections 
for the whole acceptance with $\erel$ integrated up to
5~MeV are 1790~$\pm$~20(stat.)~$\pm$~110(syst.)~mb for the Pb target,
and 93.3~$\pm$~0.8(stat.)~$^{+5.6}_{-10.3}~$(syst.)~mb for the C target 
(see the first column of Table~\ref{tab_cross}).
Here, the systematic uncertainty comes mainly from that
in the neutron detection efficiency, which affects solely the absolute
normalization of the spectrum. A minor contribution to the uncertainty
is due to the target excitation
and due to the events decaying to the $^{10}$Be excited states,
which can be significant for the carbon target data.
These contributions have been 
estimated and subtracted 
using the Q-value spectrum reconstructed from all the 
four momentum vectors of $^{11}$Be, $^{10}$Be, and the neutron. 
The events excluded with this procedure were about 4~\% and 19~\%
of the total events for the Pb and C targets, respectively.

The substantially larger cross section for the Pb target over the C target
is a clear indication of the dominance of the Coulomb breakup 
for the Pb target.
The current relative energy spectrum observed for the Pb target
is consistent with our previous experiment~\cite{NAKA94}. 
The absolute value in the current experiment is about 17\% smaller 
in the central values. This discrepancy of
the central value is within the
systematic uncertainty (of about 20\%) of the absolute value in the previous
experiment. 
The current Pb spectrum is consistent with the GSI data~\cite{PALI03} 
if one takes into consideration the different virtual 
photon spectra at the two different
incident energies. The two peaks observed
for the current carbon target data were
not observed in the GSI data at 520~MeV/nucleon~\cite{PALI03}. 
This fact may be attributed to the different contribution
of the inelastic scattering to these states for different incident energies.
The eikonal calculation in Ref.~\cite{HENC96} suggests
that the diffractive breakup, which contains the inelastic
scattering to discrete states, is expected to have 
a factor of 3--5 larger cross section 
below 100~MeV/nucleon than at high energies. 
In fact, the current cross section 
is about 3 times larger than the value of 32.6(1.6) mb
reported in Ref.~\cite{PALI03}. Due to the smaller cross section at higher
energies, the peaks might not have been statistically significant in the
GSI experiment. The larger cross section and the better energy 
resolution at intermediate energies compared to higher energies 
may be better suited for spectroscopic studies of such 
discrete unbound resonance states.

It should be noted that the 
$1n$ removal cross section in the $1n$ coincidence measurement with
the detectors placed in the forward direction adopted in the current and
in the quoted GSI experiments
corresponds to the diffractive breakup cross section, while the other
component of the reaction, the 1$n$ knockout process, is  
out of the acceptances. Taking into account the fact that
the diffractive breakup cross section is expected to be 
about half of the total $1n$ removal cross section 
at intermediate energies~\cite{HENC96}, we find that 
the current cross section is consistent with 
the value of 259(39)~mb obtained for the $^{11}$Be + $^{9}$Be 
reaction with no coincidence with the neutron at a similar incident energy
of 60~MeV/nucleon~\cite{AUMA00}.

\begin{table}[t]
\begin{center}
\begin{tabular}{cccc} \hline\hline 
Target     &$\sigma$ (mb)&    $\sigma_{\rm E1}$(mb) &   
$\sigma_{\rm NFCB}$ (mb)  \\ \hline
Pb         & 1790$\pm$20(stat.)$\pm$110(syst.) & 1510$\pm$92 & 280$\pm$20 \\
C          &  
93.3$\pm$0.8(stat.)$^{+5.6}_{-10.3}$(syst.)& 12.5$^{+0.8}_{-1.4}$ &
 80.8$\pm$0.8\\ 
\hline\hline
\end{tabular}
\caption{Cross sections of $^{11}$Be~$\rightarrow$~$^{10}$Be~+~$n$ 
on Pb and C targets for $\erel\leq$ 5~MeV. 
The cross sections calculated for the
pure E1 Coulomb breakup, 
and the subtracted cross sections (NFCB: Non-First Order E1 
Coulomb Breakup) are
also listed. Since the systematic uncertainty in $\sigma$ is for the absolute
normalization, the calculated $\sigma_{\rm E1}$ reflects mainly this
uncertainty, while the uncertainty 
in NFCB is mainly statistical.
See also Sec.~\ref{sec:nuclhigh} 
for a discussion of the NFCB component.}
\label{tab_cross}
\end{center}
\end{table}

\subsection{Pb target data}

Here below we discuss in detail the combined results of 
angular distribution and the relative energy spectrum obtained for 
the $^{11}$Be breakup on the Pb target, used 
to extract the pure Coulomb breakup contribution.

\subsubsection{Framework of the analysis}

The Coulomb breakup cross section can be factorized into
the E1 transition part (structure part) and the reaction part.
For the E1 transition, we analyze the data 
in terms of the direct breakup mechanism, which has 
successfully explained the $\beone$ distributions for the Coulomb
breakup of one-neutron halo nuclei at low excitation 
energies~\cite{NAKA94,NAKA99,PALI03}. For the reaction part, we
use two methods: 1) the semi-classical first order perturbation theory
of the equivalent photon method~\cite{JACK75,BERT88}, 2)
the quantum mechanical DWBA/coupled-channel calculation 
ECIS~\cite{RAYN97}. 
In the case of the equivalent photon method
the double differential cross section
can be given as, 
\begin{equation}
\dsdomegade = \frac{16\pi^3}{9\hbar c}\dndomega\dbderel,
\label{eq-vphoton}
\end{equation}
where $\neone$ stands for the number of virtual photons with photon
energy $\ex$ and scattering angle $\theta$.
$\beone$ is the reduced transition probability
for an E1 excitation. The photon number $\neone$ represents
the reaction part, and $\beone$ represents the structure
part. 

In the ECIS approach, we discretized the excitation energy.
For each energy bin, the $\beone$ from the structure model
was integrated to obtain the Coulomb deformation length
parameter $\delta_C$ (=$\beta R$, with deformation parameter $\beta$
and nuclear radius $R$), which was then
used as an input of the ECIS code to obtain the angular distributions.
The reaction part is also independent of 
the $\beone$ spectrum in this quantum mechanical approach. 

\subsubsection{Direct breakup mechanism}

In the direct breakup 
mechanism~\cite{NAKA94,OTSU94,MENG97,BERT92,BAUR92}, 
the $\beone$ distribution contained in Eq.~(\ref{eq-vphoton}),
is described simply by the following matrix element
\begin{eqnarray}
        \dbderel &=& 
      \mid \langle{\bf q} \mid \frac{Ze}{A}rY^1_m \mid \Phi({\bf r})\rangle 
         \mid^2 .
           \label{be1}
\end{eqnarray}

The wave function for $^{11}$Be in the ground state, $\Phi({\bf r})$,
is represented by the product of the
radial part $R(r)$ and the angular part of the single valence neutron.
The E1 operator involves $r$,
the relative distance between the core and 
valence neutron. The final state $\langle{\bf q}\mid$ describes a 
neutron in the
continuum. The matrix element represents approximately a Fourier
transform of $rR(r)$. In fact, it is an exact Fourier transform if one
neglects the interaction in the continuum and the final state
$\langle{\bf q}\mid$ is assumed to be a simple plane wave. The $\beone$ 
spectrum at low excitation energy (small ${\bf q}$) is therefore an 
amplified image of the density
distribution for large $r$, {\it i.e.}, 
the halo distribution. In other words,
the E1 amplitude at low relative energies is proportional to the asymptotic
normalization coefficient of the halo wave function.
Namely, the Coulomb dissociation probes exclusively the halo part
of the wave function. 

One can then relate the $\beone$ amplitude 
to the spectroscopic factor of the $^{11}$Be single particle state
as described in Ref.~\cite{NAKA94,NAKA99,DATT03,PALI03}.
The wave function of the ground state of $^{11}$Be can be described as,
\begin{eqnarray}
\Phi(r) &=&  \alpha\mid\shalfzero\rangle+ \nonumber\\ 
        &&    \beta\mid\dhalftwo\rangle+ ...,
\end{eqnarray}
where $\alpha^2$ and $\beta^2$ 
represent the spectroscopic factor for each configuration in the term 
expansion.
The first term is the halo configuration since the $s$-wave
valence neutron has no centrifugal barrier and, combined with the very
low binding energy, represents the halo tail.
Hence, the $\beone$ distribution at low excitation energy is sensitive only
to the first term of the wave function, as in,
\begin{equation}
\dbderel 
\propto 
\alpha^2\mid \langle{\bf q} \mid \frac{Ze}{A}rY^1_m \mid \shalfzero\rangle 
\mid^2 .
\label{direct1}
\end{equation}
The comparison of the E1 amplitude (or differential cross section)
to the theoretical expectation thus leads to 
the determination of the
spectroscopic factor for the halo state as was successfully shown
in the previous experiments on $^{11}$Be~\cite{NAKA94,MENG97,PALI03}, 
$^{15}$C~\cite{DATT03}, and $^{19}$C~\cite{NAKA99}.

\subsubsection{Angular distribution}

\begin{figure}[t]
\includegraphics[width=75mm]{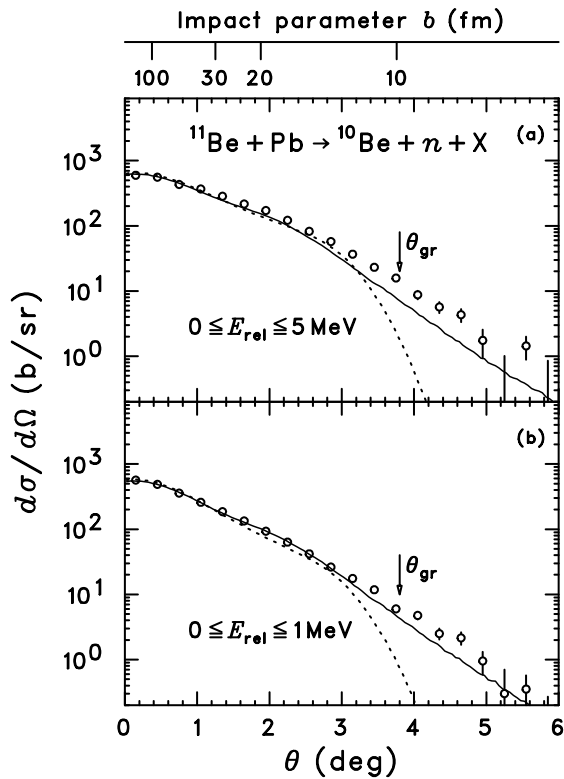}
\caption{\label{angle_pb} Angular distributions of the 
$^{10}$Be~+~$n$ c.m. system scattered by the Pb target 
for the $\erel$ ranges of 0~$\leq~\erel~\leq$~5~MeV (a)
and 0~$\leq~\erel~\leq$~1~MeV (b). 
The solid (dotted) curve shows the calculated
results with the ECIS code (equivalent photon method).
The arrows show the grazing angle $\theta_{\rm gr}$(=3.8$^\circ$).}
\end{figure}

Figure~\ref{angle_pb} shows 
the cross sections plotted as a function of
the scattering angle $\theta$ of the $^{10}$Be~+~$n$ c.m. system
of the $^{11}$Be breakup on the Pb target.
Here the angular distributions 
are shown for the two $\erel$ energy
regions of 
0~$\leq \erel\leq$~5~MeV (a) and 
0~$\leq \erel\leq$~1~MeV (b).

These angular distributions are compared to the calculations 
performed with the equivalent photon method and with the DWBA
method (ECIS).
For both cases, $\beone$ has been calculated according to 
Eq.~(\ref{direct1}), with the halo wave function obtained
using a potential model based on a Woods Saxon potential with parameters
$r_0$~=~1.236~fm and $a$~=~0.62~fm. The experimental value of the
binding energy $S_n$ was used to determine the well-depth parameter.
The final state, distorted wave function in the continuum 
$\langle{\bf q}\mid$ was calculated using the same interaction potential. 

In the equivalent photon method the angular distributions
in Fig.~\ref{angle_pb} have been obtained  by integrating the 
$\beone$ distribution over the relative energies
and by folding with the experimental resolutions.
The cutoff impact parameter for the calculation of the photon 
number adopted there is 12.3~fm, as given in Ref.~\cite{NAKA94}.
For the ECIS calculation, we assume a pure E1 Coulomb excitation 
with the optical model potential parameters determined
by fitting the $^{17}$O~+~Pb elastic scattering data at 
84~MeV/nucleon~\cite{BARR88}. The values of the parameters adopted are 
given in Table~\ref{tab_opt}.

The normalization to the experimental data has been obtained by
matching the amplitude at the most forward angles for 
the 0~$\leq~\erel\leq$~1~MeV 
data (see Fig.~\ref{angle_pb}(b)). Note that the normalization obtained 
in the analysis shown in Fig.~\ref{angle_pb}(b) (0~$\leq~\erel~\leq$~1~MeV) 
can reproduce the normalization 
used for Fig.~\ref{angle_pb}(a) (0~$\leq~\erel~\leq$~5~MeV) 
as well. The resulting $\alpha^2$ turned out to be 0.72 when using the ECIS 
technique and 0.69 when applying the equivalent photon method. 
The final determination on the spectroscopic factor and the 
discussion on its uncertainty will be presented with the relative 
energy spectrum below.

Shown in Fig.~\ref{angle_pb} 
are the angular distributions characterized by 
a forward peak and a sharp fall off, which 
can be easily understood in the semi-classical picture. 
When the Coulomb excitation occurs with a low-energy
virtual photon as in the case of a loosely bound
nucleus, the virtual photon is absorbed by the nucleus
at a large impact parameter, or small
scattering angles. In fact,
the impact parameter $b$ is related to the scattering angle
$\theta$ by $b=a\cot(\theta/2)\simeq 2a/\theta$ in this semi-classical
treatment, where $a$ stands for half the distance of the closest
approach. The impact parameter axis in this relation 
is shown on the top of Fig.~\ref{angle_pb}. One can see that 
even events at large impact parameters of
more than 100~fm can contribute to the Coulomb breakup. 

As for the comparison of the distributions, an overall agreement of
the calculation with the experimental data has been obtained at
forward angles for both energy regions, indicating the dominance of the
E1 Coulomb breakup with the direct breakup mechanism.
Between the two models, the quantum mechanical calculation
gives a better agreement for a wider range of the angular distribution.
This may be due to the fact that the
ECIS code incorporates the quantum mechanical trajectory on
the optical potential and the extended charge distribution of the target. 

In spite of the successful explanation of the data
at forward angles, there still remains a deviation at large scattering angles,
in particular beyond the grazing angle $\theta_{\rm gr}$(=3.8$^\circ$).
This can be attributed to a larger nuclear contribution and/or 
higher order effects at these angles. Even for the ECIS calculation,
a slight deviation remains for the angles above
1.3 degrees for $\erel\leq$~5MeV and above 2.8 degrees for $\erel\leq$~1~MeV.
In turn, the selection of the data 
at forward angles is proved to be very effective 
to extract the almost pure E1 Coulomb
breakup component. 
This is also supported by 
an elaborate theoretical calculation which included all higher
order effects in Coulomb and nuclear excitations~\cite{TYPE01b}.
There, the prediction has been that the pure E1 Coulomb breakup 
occurs within about one half of the grazing angle.

\begin{table*}[th]
\begin{center}
\begin{tabular}{cccccccc} \hline\hline 
Original Reaction    & Energy/nucleon 
         & $V$  & $r_v$ & $a_v$ & $W$ & $r_w$ & $a_w$ \\
& (MeV)  & (MeV) & (fm) & (fm) & (MeV) & (fm) & (fm)
\\ \hline
$^{17}$O+$^{208}$Pb~\cite{BARR88} 
& 84 & 50 & 1.067 & 0.8 & 57.9  & 1.067 & 0.8 \\ \hline
$^{11}$Be+$^{12}$C (set a)~\cite{JOHN97} 
& 48 & 155.9 & 0.632 & 0.994 & 92.66  & 0.593 & 1.042 \\ 
$^{12}$C+$^{12}$C  (set b)~\cite{BUEN84} 
& 84 & 120 & 0.71 & 0.84 & 34.02 & 0.96 & 0.69 \\
\hline\hline
\end{tabular}
\caption{Optical potential parameters used for the ECIS
calculation of the $^{11}$Be + Pb reaction at 
69 MeV/nucleon (1st row)~\cite{BARR88}, 
and those for the $^{11}$Be + C reaction
at 67 MeV/nucleon (2nd and 3rd rows)~\cite{JOHN97,BUEN84}.}
\label{tab_opt}
\end{center}
\end{table*}

\subsubsection{Relative energy spectra}
\label{sec:erel}

In Fig.~\ref{erel}(a), the relative energy spectrum selected for the forward
angles ($\theta\leq$~1.3$^\circ$) is 
compared with the pure E1 breakup of the ECIS calculation (solid curve) and 
that of the equivalent photon method (dotted curve). Since these two
calculations give almost the same results, 
the dotted curve is hardly appearing in the figure.
The angle of 1.3~degrees corresponds to 30~fm in the semi-classical framework.
This angle-selected spectrum agrees perfectly with the calculation and 
shows that the selection of the forward angular region
is, indeed, very effective to extract the E1 Coulomb breakup component. 
The spectroscopic factor for the halo ground state is thus deduced
to be 0.72~$\pm$~0.04 (ECIS), and 0.69~$\pm$~0.04 (equivalent photon
method),
which are consistent with each other. 
These values agree well with those obtained from 
the angular distribution analysis.
The extracted spectroscopic factors
are listed in Table~\ref{tab_spec} and are shown there in comparison 
with those obtained in other breakup experiments
and other experiments using different reactions.

\begin{table}[t]
\begin{center}
\begin{tabular}{cccc} \hline\hline 
Reaction    & E/A (MeV) & References & $\alpha^2$ \\ \hline
Coulomb Breakup &     69         & Present  & 0.72~$\pm$~0.04 (QM) \\
                &            &              & 0.69~$\pm$~0.04 (SC) \\          
            & 72 & \cite{NAKA94,MENG97} & 0.80~$\pm$~0.20 (SC) \\
            & 520 & \cite{PALI03}       & 0.61~$\pm$~0.05 (SC) \\
\hline
Transfer Reaction & 12.5 & \cite{ZWIE79}& 0.77 \\
                  &      & \cite{TIMO99}& 0.60,0.36 \\
                  &  35.3  & \cite{WINF01} &  0.67-0.80 \\ \hline
Knockout Reaction & 60  & \cite{AUMA00} &  0.87~$\pm$~0.13 
\\ 
                  &     &  \cite{ESBE01} & 0.79~$\pm$~0.12 \\
\hline\hline
\end{tabular}
\caption{Comparison of spectroscopic factors obtained from different
reactions. For the Coulomb breakup, 
QM (Quantum mechanical) stands for the
ECIS analysis, while SC (Semi-classical) stands for the semi-classical
equivalent photon method. For transfer reactions, Ref.~\cite{TIMO99}
is a reanalysis of the experiment of Ref.~\cite{ZWIE79}. 
For the knockout reaction, Ref.~\cite{ESBE01} is a reanalysis of
the experiment of Ref.~\cite{AUMA00} with a corrected eikonal model.}
\label{tab_spec}
\end{center}
\end{table}

The spectroscopic factor ($\alpha^2$=0.72) extracted from the
data with the restricted angular range is then used to 
calculate the spectrum for the whole acceptance (using the ECIS method).
The result of this calculation is shown in Fig.~\ref{erel}(a) where
the overall agreement with the experimental data is evident, although
a deviation at $\erel\sim 1-2$~MeV can be observed.
This deviation may be
attributed to a nuclear contribution and/or to higher order effects
in the electromagnetic excitation process.
The difference between the calculation and the data provides a measure of
these effects, as will be discussed in section~\ref{sec:nuclhigh}.
The dotted curve obtained with the equivalent photon method 
with $\alpha^2$~=~0.69 is also shown in comparison with the 
data. In this case the deviation is larger due to the impact parameter cut
as can be seen in the angular distributions of Fig.~\ref{angle_pb}. 
The integrated cross section for the pure E1 breakup calculation (ECIS)
and the residual cross section are listed in Table~\ref{tab_cross}.

The integrated $\beone$ obtained from the data selected for the
forward angles amounts to 1.05~$\pm$~0.06~e$^2$ fm$^2$ 
corresponding to 3.29~$\pm$~0.19~W.u for $\ex~\leq$~4 MeV. 
This can be compared to the E1 non-energy weighted cluster sum rule
as proposed in Ref.~\cite{ESBE92}
\begin{equation}
\beone = 
 (3/4\pi)(Ze/A)^2 \langle\, r^2\,\rangle,
\end{equation}
where $\langle\, r^2\,\rangle$ represents the mean square distance
between the valence neutron and the core, and $Z$ and $A$ 
represent the atomic and mass numbers of $^{11}$Be in the present case.
From the sum up to $\ex$~=~4~MeV, we obtain 
$\sqrt{\langle\, r^2\,\rangle}$ = 5.77~$\pm$~0.16 fm 
for the halo neutron, which is consistent with the value of
5.7~$\pm$~0.4 fm obtained from the GSI experiment.

The energy-weighted E1 sum rule (TRK sum rule) can be written as,
\begin{equation}
\int_0^{\infty} \ex B(E1;\ex)\; d\ex
     = \frac{9}{4\pi}\frac{\hbar^2 e^2}{2m}\frac{NZ}{A}.
\end{equation}
The TRK E1 sum for $^{11}$Be is 38.1~e$^2$fm$^2$MeV, while 
the observed strength
(from the one-neutron decay threshold energy
to $\ex$~=~4~MeV) amounts to
1.52~$\pm$~0.22~e$^2$fm$^2$MeV, which is only 4.0(5)\% of the 
expected total TRK sum.
In the present case, however, the
cluster sum rule is more appropriate for a comparison with the
experimental result.
The cluster E1 sum (molecular sum) $S_1$
is defined by subtracting the contribution of the core internal 
motion from the total TRK sum as in~\cite{ALHA82},
\begin{equation}
{\rm S_1} = \frac{9}{4\pi}\frac{\hbar^2 e^2}{2m}
          \left(\frac{NZ}{A} - \frac{N_c Z_c}{A_c} \right),
\end{equation}
where suffix '$c$' represents core-related quantities.
The observed sum exhausts 70~$\pm$~10~\% of the cluster sum of
2.18~~e$^2$fm$^2$MeV for the $^{10}$Be$-n$ motion. 
It should be noted that this value agrees with 
the spectroscopic factor for the halo state.
The cluster sum may provide an alternative way 
of extracting the spectroscopic factor for the halo state.

\subsection{C target data}

The relative energy spectra for $^{11}$Be~+~C at 67 MeV/nucleon shown
in Fig.~\ref{erel}(b) have been investigated 
in combination with angular distributions.
We first describe here the results on the two observed peaks by
showing their angular distributions and compare the results with
shell model calculations. We then further investigate the
angular distribution to extract information on the reaction mechanism
of the breakup with a light target.

\subsubsection{1.78 MeV and 3.41 MeV states}

\begin{figure}[t]
\includegraphics[width=85mm]{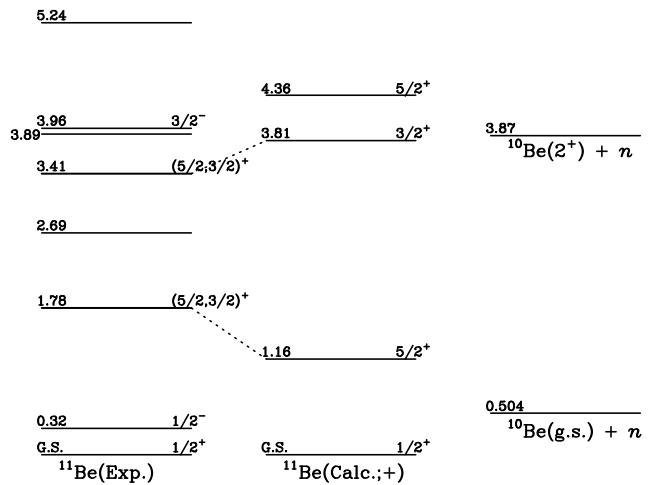}
\caption{\label{level} Experimental and theoretical energy levels of 
$^{11}$Be. The experimental spectrum is from Ref.~\cite{AJZE90}, and
includes the present spin-parity assignments for 
the 1.78~MeV and 3.41~MeV
states. The theoretical spectrum for the positive parity states of
 $^{11}$Be was obtained by shell model calculations
(OXBASH) with the WBT interactions.
The energy levels of $^{10}$Be(g.s.) + $n$ 
and the first excited state of $^{10}$Be+ $n$ are also shown.}
\end{figure}

In the relative energy spectrum for the $^{11}$Be breakup on the C target, 
two resonance peaks centered at $\erel$~=~1.29~MeV ($\ex$~=~1.79~MeV) and 
$\erel$~=~2.88~MeV ($\ex$~=~3.38~MeV) have been observed with significant
strengths embedded on the continuum component. These states
correspond to the known states in $^{11}$Be
at $\ex$=~1.78~MeV (2nd excited state) and at 
$\ex$=~3.41~MeV, as shown in Fig.~\ref{level}~\cite{AJZE90}.
So far, these states have been identified by transfer reactions on
$^{10}$Be~\cite{ZWIE79} and $^{9}$Be~\cite{LIU90}. The 1.78 MeV state has been
assigned to be a $J^{\pi}=(5/2,3/2)^+$.
As for the 3.41 MeV state, the spin-parity assignment has been
more controversial, with positive parity
$J^{\pi}=(1/2,3/2,5/2)^+$~\cite{AJZE90}
and negative parity assignments $J^{\pi}=3/2^-$~\cite{LIU90}.

\begin{figure}[th]
\includegraphics[width=80mm]{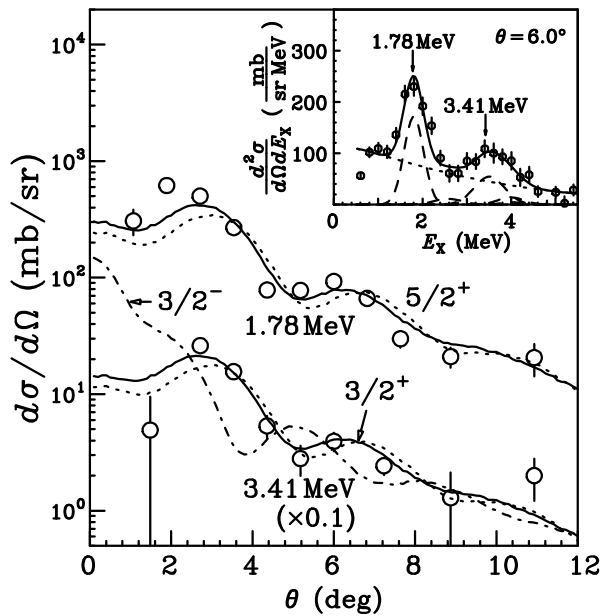}
\caption{\label{angle_c} Angular distributions for $\ex$~=~1.78 MeV and
 3.41MeV states. The solid and dotted curves are 
obtained by ECIS calculations for
the $L$=2 transitions to the 1.78~MeV with $J^\pi$=5/2$^+$
and the 3.41~MeV state with $J^\pi$=3/2$^+$, where the
optical potential parameters are
from the $^{11}$Be+$^{12}$C reaction (solid curves, set a),
and the $^{12}$C+$^{12}$C reaction (dotted curves, set b). 
The calculation assuming $L=1$
is shown by the dot-dashed curve for the 3.41 MeV state excitation.
In the inset an example of
the fitting of the excitation energy spectrum used to extract the
cross section at a fixed angle is shown.}
\end{figure}

The angular distributions for the excitation of these states have 
been obtained by fitting the relative energy spectrum for each
$\theta$ bin.
The fitting function consists of Gaussians corresponding to 
known discrete states up to $\ex$~=~5.24~MeV
plus arbitrary exponential and polynomial functions 
for the representation of the
continuum background component. One example of the fitting result
is shown in the inset of Fig.~\ref{angle_c}. 
We find that only the 1.78~MeV and 3.41~MeV states
have significant cross sections amongst the known states.

The angular distributions thus obtained are shown in
Fig.~\ref{angle_c}.
The integrated cross sections up to $\theta$~=~12$^\circ$
amount to 10.7~$\pm$~2.1~mb and 5.9~$\pm$~1.2~mb respectively
for the 1.78~MeV and 3.41~MeV states. 
A systematic uncertainty of about 20\% arising from
the ambiguity of the choice of the continuum background function
is incorporated in the quoted uncertainties.
We have compared these data
with the DWBA calculation (ECIS code)
using a standard vibrational model.
Both angular distributions follow the
patterns characteristic of $L$=2 transition. 
In the ECIS calculations, we have adopted the $5/2^+$ and $3/2^+$ 
assignments respectively for 1.78~MeV and 3.41~MeV to reproduce
the $L$=2 transitions, although
the choice of either the $3/2^+$ or $5/2^+$ assignment does not
modify the angular pattern.
The assignment in the ECIS calculation
reflects the agreement with the level 
order in a shell model calculation for the positive parity 
states as described below (see Fig.~\ref{level}).
Two different
optical potential parameter sets (a)~\cite{JOHN97} and (b)~\cite{BUEN84},
given in Table~\ref{tab_opt}, are adopted for extracting the differential
cross sections.
The calculations are in good agreement with the data
for both parameter sets.
As for the 3.41 MeV state, 
$L = 1$ assignment ($J^\pi$=3/2$^-$) is clearly
excluded, as can be seen in Fig.~\ref{angle_c}.

The deformation lengths $\delta(=\beta R$) can be 
obtained from the DWBA analysis as well. The results 
are listed in Table~\ref{tab_delta}.
These deformation lengths as well as the experimental energies
are compared to shell model calculation
for the positive parity states in $^{11}$Be 
in the $p-sd$ model space with the 
WBT effective interactions~\cite{OXBA86,WARB92}.
The comparison of the experimental and calculated energy levels
is shown in Fig.~\ref{level}.
The deformation lengths can be obtained by 
introducing Bernstein's prescription~\cite{BERN81},
\begin{equation}
\delta = C \frac{4\pi}{3eR} \frac{b_p M_p + b_n M_n}{b_p Z + b_n N},
\end{equation}
where $M_p$ and $M_n$ are proton and neutron multipole matrix elements.
The parameters $b_p$ and $b_n$ represent the interaction strengths
of the probe particle, respectively 
for protons and neutrons. 
We adopt $b_p=b_n=1$ since $^{12}$C is a $T=0$ probe as in the case of
$\alpha$ particles,
where the same parameters are used.
In the vibrational model, the factor $C$ is given by
\begin{equation}
C = \sqrt{\frac{5}{2I_f + 1} }, \nonumber
\end{equation}
where $I_f$ represents the nuclear spin of the final state.
With these prescriptions, shell model calculations provide 
the matrix elements $M_p$ and $M_n$ from which the theoretical
deformation lengths have been deduced. The results are
presented in Table~\ref{tab_delta}. 
In this calculation, we have adopted the conventional effective charges of 
$e_p=1.3e$ and $e_n=0.5e$, which are commonly used in the 
$sd$-shell region~\cite{BROW88}. The calculated deformation lengths 
are consistent with the experimentally obtained values, with a
better agreement with the results obtained using
the optical potential parameter set (a).

The reasonably good agreement both for the level
energies and the transition strengths shows that the shell model
describes rather well these positive-parity states with spin
assignments of 5/2$^+$ and 3/2$^+$ respectively for 1.78~MeV and
3.41~MeV states.
The main shell model configuration for the 1.78~MeV state 
is $\dhalfzero$.
On the other hand the 3.41 MeV state has only a small fraction
of 1$d_{3/2}$ single particle component, and a larger contribution 
of the $\shalftwo$ configuration.
It should be noted here that even if the 3.41 MeV state 
has a large component with the $^{10}$Be(2$_1^+$) excited core,
this configuration decays into $^{10}$Be(g.s.) since the decay
into $^{10}$Be(2$_1^+$) is energetically forbidden (see Fig.~\ref{level}). 

We note here that the inelastic scattering on a $T$=0 target as
in $^{12}$C rather favors the $L$~=~2 excitation because this reaction
yields in principle, isoscalar excitations. 
This may be the reason for the observations of
these positive parity states in the present experiment.
The inelastic scattering on a different target, such as a proton,
would be very interesting to study since this would excite states with
different $\jpi$.
A comparison of the transition strengths for the observed 1.78~MeV and
3.41~MeV states probed
by a different target would be also 
interesting since this would lead an independent determination
of $M_n$ and $M_p$, thereby enabling the extraction of
different neutron and proton deformations.

\begin{table}[t]
\begin{center}
\begin{tabular}{cccc} \hline\hline 
State &   $\jpi$     &     & $\delta$ (fm) \\ \hline
1.78~MeV & 5/2$^+$      & Exp. (set a) & 1.27~$\pm$~0.25 \\
         &               & Exp. (set b) & 1.16~$\pm$~0.23 \\
         &               & Shell Model                & 1.23 \\ \hline
3.41~MeV & 3/2$^+$      & Exp. (set a) & 1.42~$\pm$~0.28 \\
         &               & Exp. (set b) & 1.02~$\pm$~0.20 \\
         &               & Shell Model                & 1.36 \\ \hline\hline
\end{tabular}
\caption{Deformation lengths obtained for the 1.78~MeV and
 3.41~MeV states with the 
two different optical potential parameter sets (set a and set b), 
compared with shell model calculations with
Bernstein's prescription~\cite{BERN81}. 
The spin-parity assignments of $\jpi=5/2^+$ and $3/2^+$ respectively for the
1.78~MeV and 3.41~MeV states have been adopted.}
\label{tab_delta}
\end{center}
\end{table}

\subsubsection{Angular distribution and E1 Coulomb component}
\label{sec:MDA}

\begin{figure}[t]
\includegraphics[width=85mm]{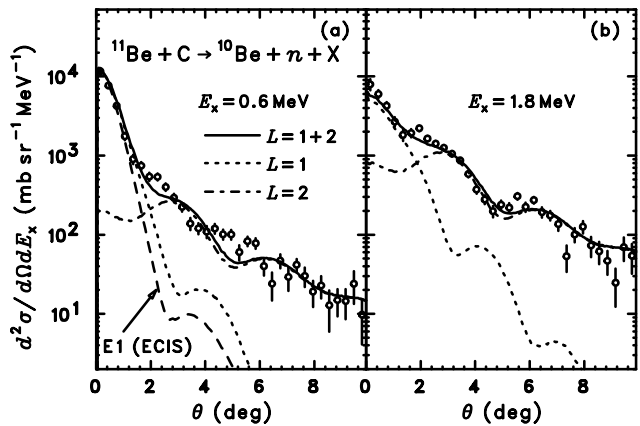}
\caption{\label{md} 
Angular distributions
of $^{11}$Be on the C target for (a) 
0.5~MeV~$\leq~\ex~\leq$~0.7~MeV corresponding to the structureless 
continuum, and for (b) 1.7~MeV~$\leq~\ex~\leq$~1.9~MeV 
corresponding to the region containing the 1.78~MeV
state. The dotted, dot-dashed, and solid lines represent the component
of $L$=1, $L$=2, and their sum, respectively. The calculation 
of the pure E1 direct breakup with $\alpha^2$~=~0.72 is also
shown by a dashed curve for (a).}
\end{figure}

A further investigation of the reaction mechanism of the $^{11}$Be
breakup on the C target based on the analysis of angular distributions
has been performed. These are shown in Fig.~\ref{md} for a pure
continuum region just above the neutron threshold (a), and in the
region containing the 1.78~MeV state (b). We find that the angular
distributions are characterized by a strong peak at very forward angles
and a diffraction pattern at larger angles. These angular distributions
are compared to ECIS calculations 
with a restriction of excitation multipoles to $L$=1 and $L$=2.
The curves labeled $L$=1 in Fig.~\ref{md} consists of calculation
performed using E1 Coulomb excitation and the isoscalar component of the
nuclear excitation~\cite{HARA81} while for the $L$=2 case a vibrational
model for Coulomb and nuclear excitations has been adopted. 
An overall agreement of
the data with this decomposition is obtained. In Fig.~\ref{md}(b), we
see a dominance of the $L$=2 diffraction pattern, as expected. The
remaining small deviations may be attributed to contributions from the
$L$=1 isovector nuclear excitation, higher multiple excitations, or
from the events decaying into the $^{10}$Be excited states. Also shown
in Fig.~\ref{md}(a) is the result of the pure E1 Coulomb calculation
with the direct breakup mechanism 
assuming a spectroscopic factor $\alpha^2$=0.72.

As a main result of this analysis we find
that the notable peak at the forward angles
is reproduced perfectly by the Coulomb breakup.
We see the strong forward peak even in the Fig.~\ref{md}(b)
which shows the angular distribution 
for the excitation of the state at 1.78~MeV. 
This result led us to compare the relative energy
spectrum for the C target at the selected angular
range $\zerozerofive$.  As shown in Fig.~\ref{erel}(b), 
we have obtained indeed an excellent agreement with the E1 direct
Coulomb breakup model
even for the breakup with a light target such as carbon.

The present work on a light target 
has demonstrated that the invariant mass spectroscopy
in combination with the angular distribution analysis offers
a very useful spectroscopic tool for a loosely-bound nucleus,
where unbound states are easily populated.
In the present analysis, collective models have been adopted 
in the calculations for simplicity.
It would be interesting to compare more elaborate microscopic 
reaction models to the present results.
For instance, the Continuum-Discretized Coupled Channel (CDCC) method
can be one of these powerful tools. 
The present results may provide a good test of
such reaction theories of the breakup of loosely bound nuclei.

\subsection{Remarks on the nuclear contribution and higher order effects}
\label{sec:nuclhigh}

In this section, we make an estimation of the nuclear and high
multipolarities in the $^{11}$Be~+~Pb data.
Practically, this estimation can be used 
to test the scaling between the nuclear breakup
component in the breakup on Pb and that on C, which has been
conventionally used for the estimation of the nuclear contribution.  
There, the Coulomb breakup spectrum has been extracted
by subtracting the nuclear contribution estimated by the spectrum with 
a light target data as in,
\begin{equation}
\frac{d\sigma_{\rm CD}}{d\erel} = 
\frac{d\sigma}{d\erel}({\rm Pb}) - \Gamma\frac{d\sigma}{d\erel}({\rm C}),
\label{scale}
\end{equation}
where suffix ``CD'' stands for Coulomb dissociation, and $\Gamma$
is the scaling factor. This scaling method assumes that the breakup
cross section is an incoherent sum of the Coulomb and nuclear
diffractive dissociation. This method may be 
important for an experiment with smaller yields
where the angular distribution as in the present work is hardly obtained.
In the previous breakup experiment of $^{11}$Be on Pb at 
72 MeV/nucleon~\cite{NAKA94}, we adopted 
the $\Gamma$ parameter to be the ratio of the sum of the radii of 
the target and the projectile, which is 1.8. This is based on a
simple geometrical argument that the nuclear excitation is a 
peripheral phenomenon. 
On the other hand, Ref.\cite{PALI03} extracted a larger value of 
$\Gamma=$~5.4 based on the eikonal calculation. 
In theoretical works, 
this ratio varies very much: a simple Serber model~\cite{SERB47} that has 
a $A^{1/3}$ target mass dependence gives $\Gamma$~=~2.6 for Pb/C,
while the model in Ref.~\cite{NAGA01}, where $A^1$ target mass dependence
is suggested, provides $\Gamma$~=~17.

\begin{figure}[ht]
\includegraphics[width=80mm]{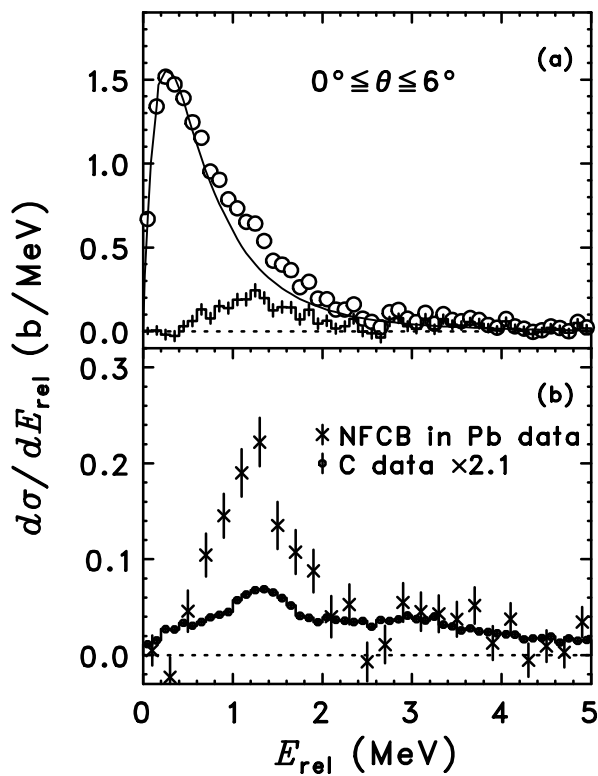}
\caption{\label{erel_nucl_c} 
(a)~The $\erel$ spectrum for the Pb target (open circles) for the
whole acceptance ($\zerosix$) is compared with
the pure 1st-order E1 Coulomb breakup calculation with 
$\alpha^2$~=~0.72(solid line).
The difference between them (histogram) represents the
non first-order Coulomb breakup (NFCB) component.
(b)~The NFCB component is compared 
to the C target data with the scaling factor $\Gamma$~=~2.1.}
\end{figure}

In the present work, we have extracted the Coulomb breakup contribution
independently from this scaling factor by using the angular distribution. 
Therefore, the extracted pure Coulomb component can be used to estimate
the scaling factor $\Gamma$.
Figure~\ref{erel_nucl_c} displays the $\erel$ spectrum for the Pb target
and for the whole acceptance ($\zerosix$) in comparison with 
the extracted E1 pure Coulomb direct breakup component 
(ECIS calculation, solid line)
with the spectroscopic factor $\alpha^2$=0.72.
The difference between the data and the estimated Coulomb contribution
shown by the histogram provides nuclear contribution and/or
higher order effects of Coulomb breakup. Since 
this difference represents the remaining breakup contribution
after subtracting first-order Coulomb breakup, 
we call here this component NFCB, the non first-order Coulomb breakup.
The NFCB component amounts to 280~$\pm$~20~mb 
which is 15.6~\% of the total breakup cross
section as shown in Table~\ref{tab_cross}.
Since we had an evidence for a Coulomb breakup component with the 
carbon target data, the NFCB in the carbon target was 
also extracted and the result is
80.8~$\pm$~0.8~mb, also given in the same table.

The NFCB spectrum is compared with that
for the C target in Fig.~\ref{erel_nucl_c}(b).
The C target data is scaled to match the integrated cross section
for $\erel\geq$2~MeV, where a good agreement in the spectral shape
is obtained.  With this comparison, we have obtained 
$\Gamma$ = 2.1~$\pm$~0.5, 
consistent with the value of 1.8 adopted in the previous
experiment, and with the value obtained from the Serber model.  
This value is also consistent with the eikonal calculation which gives
$\Gamma$=~2.4 at this incident energy~\cite{IBRA03}. 
However we find a strong deviation at
the resonance region near the 1.78~MeV state, where $\Gamma$ is about 6.
The cause of this large deviation cannot be easily understood. We infer that
this may be either due to a strong nuclear and Coulomb interference for 
this particular resonant state. The other possible reason is
a higher order Coulomb excitation effect around this energy region. 
In any cases, the possibility
of mass dependence of $A^1$ for the nuclear breakup cross sections
in Ref.~\cite{NAGA01} can be clearly excluded.
Due to the consistency at the energy region where the structureless 
diffractive dissociation dominates, we may adopt the $\Gamma$~=~2.1(5) 
as an estimation of the nuclear contribution at this incident energy.

The value of $\Gamma$~=~2.1(5) 
is smaller than that obtained at GSI at 520~MeV/nucleon. 
According to eikonal calculation~\cite{HENC96}, 
the energy dependence of $\Gamma$
can be understood by the fact that at higher energies
the black-disk like picture is more vague by low NN cross
sections. Namely, the Serber-type picture is to be modified 
at higher energies.

\section{Summary and Conclusions}
\label{summary}

In this paper, we have investigated the Coulomb and nuclear breakup
of $^{11}$Be on Pb and C targets
at 69~MeV/nucleon and 67~MeV/nucleon, respectively.
By measuring the momentum vectors of the incoming $^{11}$Be,
of the outgoing $^{10}$Be, and the neutron in coincidence, 
we have extracted the relative energy
spectra as well as the angular distributions of the scattering of 
the $^{10}$Be~+~$n$ c.m. system on both targets. 

The obtained angular distribution of the $^{10}$Be~+~$n$ c.m on 
Pb has been found to be well described by the first-order E1 
Coulomb breakup mechanism, in particular 
for the very forward angular regions corresponding to
large impact parameters in the semi-classical point of view.
The experimental relative energy spectrum selected for $b\geq$~30 fm 
(or $\theta\leq$~1.3$^{\circ}$) is in perfect agreement with the
1st order pure E1 Coulomb calculation with the direct breakup mechanism, 
leading to a spectroscopic factor of the halo configuration $\shalfzero$
of 0.72~$\pm$~0.04. 

The E1 non-energy weighted sum,
corresponding to the integrated $\beone$ strength has been compared
to the cluster sum rule, leading to a root-mean square distance
of 5.77~$\pm$~0.16~fm for the neutron in its halo state. 
The energy weighted cluster sum rule has been applied to
the present case. The energy weighted E1 strength has been
found to be 70~$\pm$~10~\% of this sum rule. It is interesting to note 
that this value agrees with the value of $\alpha^2$=0.72(4).

We have investigated experimentally the
inelastic breakup scattering of $^{11}$Be on C target.
We have observed two peaks corresponding to the $\ex$~=~1.78~MeV and
3.41~MeV states. The angular distributions for these states show
$L$=2 excitation patterns, leading to their spin-parity
assignments of $J^{\pi}$=(3/2,5/2)$^+$. The amplitude of the
angular distribution has provided the deformation lengths for
these states as well. The energy levels and the transition densities
have been compared to shell model calculations for the low-lying
positive parity states of $^{11}$Be in the  $p-sd$ model space
with the WBT effective interaction. We have found that the energy levels
and deformation lengths are rather well reproduced. The comparison with
the shell model calculation also suggests $\jpi$~=~5/2$^+$ and 
$\jpi$~=~3/2$^+$ as preferred assignments for the
$\ex$~=~1.78~MeV and 3.41~MeV states, respectively.
The deformation lengths are also well reproduced with the shell model
calculation with Bernstein's prescription and the conventional values of
effective charges.

The angular distributions have been investigated further to
disentangle the reaction mechanism.
We have found that the $L$=1 Coulomb component is 
strong at very forward angles in contrast to the $L$=2 pattern
in the angular distribution around the 1.78 MeV resonance.
In fact, the relative energy spectrum for $\theta\leq 0.5^\circ$
is well reproduced by the pure E1 direct breakup model.
This result shows that Coulomb breakup occurs at forward angles
even on a light target such as carbon.

Finally, we have estimated the nuclear breakup and/or higher
order effects by subtracting the calculated pure E1 Coulomb
component for the Pb target.
By making a comparison of the subtracted spectrum with the C
target, the scaling factor $\Gamma$ of nuclear contribution for Pb target 
to C target is estimated.  The scaling factor needed to reproduce
the data at $\erel\geq$~2MeV has been found to be $\Gamma$~=~2.1(5), 
which is consistent with Serber-type models. 
This value is smaller than $\Gamma$=5.4 extracted from data at 
higher energies. This incident energy dependence can be qualitatively
explained in the eikonal picture.

The present work demonstrates that breakup reactions, 
both on light and heavy targets, are powerful spectroscopic tools 
for low-lying states of loosely bound nuclei where 
the excitation above the particle emission threshold is close 
to the ground state. In particular, this work shows that the combination 
of angular distribution data with the relative energy spectra 
is very effective for extracting structure information by 
disentangling Coulomb and nuclear excitations.
It can be easily foreseen that the study of the inelastic
scattering to states above the threshold with
different targets would provide complementary information
on the excitation process and on the structure of excited states
in the continuum. 
More elaborate theoretical work on breakup reactions would be desirable 
to construct the spectroscopic properties in a more microscopic way
for the future RI beam science.

\section*{Acknowledgements}

Sincere gratitude is extended to 
the accelerator staff of RIKEN
for their excellent operation of the beam delivery. 
Fruitful discussions with T.~Motobayashi, K.~Yabana, M.~Ueda,
M.~Takashina, and K. Hencken are greatly appreciated. We thank also
B.~Abu-Ibrahim, Y.~Ogawa, and Y.~Suzuki for the useful suggestions 
on the eikonal calculation code~\cite{IBRA03}. 
The present work was supported in part
by a Grant-in-Aid for Scientific Research (No.15540257) from
the Ministry of Education, Culture, Sports, Science and Technology
(MEXT, Japan).


\begin{references}
\bibitem{HANS95} P.G. Hansen, A.S. Jensen, and B. Jonson, Annu. Rev. Nucl.
Sci. {\bf 45}, 591 (1995), and references
                 therein.
\bibitem{TANI95} I. Tanihata, Prog. Part. Nucl. Phys. {\bf 35}, 505 (1995),
                 and references therein.
\bibitem{TANI85} I. Tanihata, H. Hamagaki, O. Hashimoto,
Y. Shida, N. Yoshikawa, K. Sugimoto, O. Yamakawa, 
T. Kobayashi, and N. Takahashi, Phys. Rev. Lett. {\bf 55}, 2676 (1985).
\bibitem{KOBA88} T. Kobayashi {\it et al.}, 
Phys. Rev. Lett. {\bf 60}, 2599 (1988).
\bibitem{ORR92} N. Orr {\it et al.}, Phys. Rev. Lett. {\bf 69}, 2050 (1992).
\bibitem{AUMA00} T. Aumann {\it et al.}, Phys. Rev. Lett. {\bf 84}, 35
(2000).
\bibitem{KOBA89} T. Kobayashi {\it et al.}, Phys. Lett. B {\bf 232}, 51
(1989).
\bibitem{IKED92}  K.~Ikeda et al., Nucl.~Phys. A {\bf 538}, 355c (1992).
\bibitem{NAKA94} T. Nakamura {\it et al.}, Phys. Lett. B {\bf 331},
         296 (1994).
\bibitem{NAKA99} T. Nakamura {\it et al.}, Phys. Rev. Lett. {\bf 83},
 1112 (1999).
\bibitem{AUMA99} T. Aumann {\it et al.}, Phys. Rev. C {\bf 59}, 1252 (1999).
\bibitem{IEKI93} K. Ieki {\it et al.}, Phys. Rev. Lett. {\bf 70},
         730 (1993); D. Sackett {\it et al.}, Phys. Rev. C {\bf 48}
118 (1993).
\bibitem{SHIM95} S. Shimoura {\it et al.}, Phys. Lett. B {\bf 348}, 29
(1995).
\bibitem{ZINS97} M. Zinser {\it et al.}, Nucl. Phys. A {\bf 619},
151 (1997).
\bibitem{LABI01} M. Labiche {\it et al.}, Phys. Rev. Lett. {\bf 86}, 600
(2001).
\bibitem{OTSU94} T. Otsuka {\it et al.}, Phys. Rev. C {\bf 49}, 2289R
 (1994).
\bibitem{MENG97} A. Mengoni {\it et al.},
in {\it Proceedings of the International
Symposium on Capture gamma-ray and Related topics}, edited by G.L. Molnar,
T. Belgya and Zs. Revay, (Springer, 1997), p. 416.
\bibitem{NAGA01} M.A. Nagarajan, C.H.Dasso, S.M. Lenzi, and A. Vitturi,
Phys. Lett. B {\bf 503}, 65 (2001).
\bibitem{DASS99} C.H. Dasso, S.M. Lenzi, and A. Vitturi, Phys. Rev. C
{\bf 59}, 539 (1999).
\bibitem{TYPE01a} S. Typel and G. Baur, Phys. Rev. C {\bf 64}, 024601
 (2001).
\bibitem{TYPE01b} S. Typel and R. Shyam, Phys. Rev. C {\bf 64}, 024605
 (2001).
\bibitem{BERT92} C.A. Bertulani and L.F. Canto, Nucl. Phys. A {\bf 539},
                 163 (1992).
\bibitem{BAUR92} G. Baur, C.A. Bertulani, and D.M. Kalassa, 
Nucl. Phys. A {\bf 550}, 527 (1992).
\bibitem{ESBE95} H.~Esbensen, G.F. Bertsch, and C.A. Bertulani,
 Nucl. Phys. A {\bf 581}, 107 (1995).
\bibitem{MARG02} J.~Margueron, A. Bonaccorso, and D.M. Brink, 
Nucl. Phys. A {\bf 703}, 105 (2002).
\bibitem{MARG03} J.~Margueron, A. Bonaccorso, and D.M. Brink, 
Nucl. Phys. A {\bf 720}, 337 (2003).
\bibitem{THOM01} I.J. Thompson and J.A. Tostevin, Nucl. Phys. A 
{\bf 690}, 294c (2001).
\bibitem{BAUR03} G. Baur, K. Hencken, and D. Trautmann,
 Progr. Part. Nucl. Phys. {\bf 51}, 487 (2003).
\bibitem{AUDI93} G. Audi and A.H. Wapstra, Nucl. Phys. A {\bf 565}, 1 (1993);
                 G. Audi, A.H. Wapstra and M. Dedieu, Nucl. Phys. A {\bf 565},
                 193 (1993), and references therein.
\bibitem{ANNE93} R. Anne {\it et al.}, Phys. Lett. B {\bf 304}, 55 (1993);
                 R. Anne {\it et al.}, Nucl. Phys. A {\bf 575}, 125 (1994).
\bibitem{PALI03} R. Palit {\it et al.}, Phys. Rev. C {\bf 68}, 
034318 (2003).
\bibitem{KUBO92} T. Kubo {\it et al.}, 
Nucl. Instrum. and Methods., B {\bf 70}, 309 (1992).
\bibitem{HENC96} K. Hencken, G. Bertsch, and H. Esbensen,
Phys. Rev. C {\bf 54}, 3043 (1996).
\bibitem{JACK75} J.D. Jackson, {\it Classical Electrodynamics, 2nd Edition}
(Wiley, New York 1975).
\bibitem{BERT88} C. Bertulani and G. Baur, Phys. Rep. {\bf 163},
299 (1988).
\bibitem{RAYN97} J. Raynal, Coupled channel code ECIS97, also 
Notes on ECIS94, unpublished.
\bibitem{DATT03} U. Datta Pramanik et al., Phys. Lett. B {\bf 551}, 63 (2003).
\bibitem{BARR88} J. Barrette {\it et al.}, Phys. Lett. B {\bf 209}, 182
 (1988); R. Liguori Neto {it et al.}, Nucl. Phys. A {\bf 560}, 733 (1993).
\bibitem{ZWIE79} B.~Zwieglinski, W. Benenson, R.G.H. Robertson, and
 W.R. Coker, Nucl. Phys. A {\bf 315}, 124 (1979).
\bibitem{TIMO99} N.K. Timofeyuk and R.C. Johnson, Phys. Rev. C 
{\bf 59} 1545, (1999).
\bibitem{WINF01} J.S. Winfield {\it et al.}, Nucl. Phys. A {\bf 683}, 48
(2001).
\bibitem{ESBE01} H.~Esbensen and G.F. Bertsch, Phys. Rev. C {\bf 64},
014608 (2001).
\bibitem{ESBE92} H. Esbensen and G.F.~Bertsch, Nucl. Phys. A {\bf 542},
		            310 (1992).
\bibitem{ALHA82} Y.~Alhassid, M.~Gai, and G.F.~Bertsch, Phys. Rev. Lett.
{\bf 49}, 1482 (1982).
\bibitem{AJZE90} F. Ajzenberg-Selove, Nucl. Phys. A {\bf 506}, 1 (1990).
\bibitem{LIU90} G.-B.~Liu and H.T.~Fortune, Phys. Rev. C {\bf 42}, 167
(1990).
\bibitem{JOHN97} P.~Roussel-Chomaz, private communication;
R.C.~Johnson, J.S.~Al-Khalili, and J.A.~Tostevin, Phys. Rev. Lett.,
{\bf 79}, 2771, (1997); The optical potential parameters were calculated by
 M.~Takashina, private communication.
\bibitem{BUEN84} M.~Buenard {\it et al.}, Nucl. Phys. A {\bf 424}, 313
(1982).
\bibitem{OXBA86} OXBASH, The Oxford, Buenos-Aires, Michigan State,
Shell Model Program, B.A. Brown, A. Etchegoyen, W.D.M. Rae, MSU
Cyclotron Laboratory Report number 524 (1986).
\bibitem{WARB92} E.K. Warburton and B.A. Brown, Phys. Rev. C {\bf 46},
                             923 (1992).
\bibitem{BERN81} A.M. Bernstein, V.R. Brown, and V.A. Madsen, Phys. Lett. B
{\bf 103}, 255 (1981); 
A.M. Bernstein, V.R. Brown, and V.A. Madsen, Comments Nucl. Part. Phys.
{\bf 11} 203 (1983)
\bibitem{BROW88} B.A. Brown and B.H. Wildenthal, Ann. Rev. Nucl. Part.
Sci. {\bf 38}, 29 (1988).
\bibitem{HARA81} M.N.~Harakeh and A.E.L.~Dieperink, Phys. Rev. C {\bf
 23}, 2329 (1981).
\bibitem{SERB47} R. Serber, Phys. Rev. {\bf 72}, 1008 (1947).
\bibitem{IBRA03} B.~Abu-Ibrahim, Y.~Ogawa, Y.~Suzuki, and I.~Tanihata,
Comp. Phys. Com. {\bf 151}, 369 (2003).
\end{references}
\end{document}